# Citation Analysis of Innovative ICT and Advances of Governance (2008-2017)

Shuhua Monica Liu[1], Liting Pan and Xiaowei Chen

Department of Public Administration, Fudan University, Shanghai 200433, China

**Abstract**

This paper opens by introducing the Internet Plus Government (IPG), a new government initiative emerging in the last decade. To understand benefits and challenges associated with this initiative worldwide, we conducted analyses on research articles published in the e-governance area between 2008 and 2017. Content analysis and citation analysis were performed on 2105 articles to address three questions: (1) What types of new ICT have been adopted in the IPG initiative in the past decade?  (2) How did scholars investigate interactions between the new ICTs and governance core to IPG? (3) How did the new ICTs interact and shape while also being shaped by the evolution of governance in the past decade? Our analysis suggests that IPG initiative has enriched the government information infrastructure. It presented opportunities to accumulate and use huge volume of data for better decision making and proactive government-citizen interaction. At the same time, the advance of open data, the widespread use of social media and the potential of data analytics also generated great pressure to address challenging questions and issues in the domain of e-democracy.

1.  **Introduction**

The wide use of the new generation of ICT, including the Internet of Things (IoT), cloud computing, big data, machine learning and artificial intelligence (AI), the mobile internet, among others, are reshaping values and practices of government, businesses, and society (Li, 2017; Keane, 2016). This new government initiative, coined as Internet Plus Government, represents a new governance form, which gives full play to Web 2.0 in government innovation and social development in the last decade (Li, 2017).  It aims to incorporate the depth of innovation of Web 2.0 in the modern governance process. Its goal is to promote innovation and productivity in both the government and the society. Governments adopting this movement hope to develop a more sustainable model of social development which utilizes Web 2.0 as the fundamental facility and implementation tool throughout the governance processes and activities.

---
[1] S. Liu is the corresponding author and her e-mail address is shuhualiu@fudan.edu.cn.

As the idea of Internet Plus Government is increasingly implemented in many governments around the world, new opportunities and challenges emerge. Government officials, scholars and citizens are still attempting to understand the implications of the newest wave of technology innovations. Questions related to the use of IoT, Cloud computing, Big data, Machine learning and AI largely remain to be answered, when it comes to the complicated interactions among people, new ICT innovation and the governance process.

We focus to address following questions through the analysis: (1) What types of innovative ICT have been adopted in the governance process in the Western countries while public administration continued to evolve in the past decade (2008-2017)? (2) How did scholars investigate interactions between innovative ICT and governance core to Internet Plus Government after 2008? (3)How did Innovative ICT interact and shape while also being shaped by the evolution of governance in the past decade? (4) What is the relationship between the numbers of publications using different ICT technologies and countries?

## 2. Citation Data and Analysis

Extending the previous year's dg.o special issue editorial (Kim and Zhang, 2016), this paper focuses on further analyzing refereed journal articles on interactions between innovative technology adoption and use, and governance published in leading academic databases 2008–2017. The idea behind this was that refereed journal articles not only set quality standards but also provide a filter, thus establishing the nature and scope of the ideas presented to the academic community in the last ten years.

Our preliminary database included 1203 articles that assessed technology adoption and use in the governance process since 2008 (Wiley Online Library - Journals, Web of Science Core Collection - Social Sciences Citation Index, EBSCO – Academic Search Premier, Springer Link Journal, and JSTOR). These represented roughly one fifth of articles published in the electronic governance domain during this specific time period in major journals and conference proceedings in the field of public administration, information science, computer science and engineering, and electronic government. The articles we chose to omit from the final dataset addressed a wide variety of topics, including pure technology and system design and commercial technology use that had no relevance to our chosen focus of study. 360 articles addressing the topic of the adoption, implementation, and diffusion of the newest generation of ICT in governance globally since 2008 were the basis for our content analysis.

The content analysis process comprised five steps:

First, fifty papers were randomly selected from the literature database for a pilot content analysis. Research team members[2] were asked to attach classification labels to research ideas and practice addressed in each paper. Classification labels used were selected and adapted from those previously used in public administration research by Munoz and Hernandez (2010) and information science research by Hawkins (2001).

---

[2] Our research team is composed of eight research assistants. They analyzed all the papers and collaborated with me on the coding process. Without the support of Liting Pan, Yupei Lei, Xiaowei Chen, Zhixin Cai, Anqi Li, Zheng Liu, Jiajun Wang and Di Qi, this paper will not be possible.

Second, a preliminary classification system was created integrating labels attached to each of the 50 papers. Brainstorming meetings were held to decide on the accuracy of 21 classification themes assigned to each paper before an updated classification system was created. An extensive memo book that recorded our decision criteria and guided our decisions in the pilot content analysis was also created. This memo book was refined and developed while members of the research team continued analyzing all articles.

Third, each of the 360 articles was then analyzed separately employing the newly created classification system following the same procedure using consistent, computer-based coding and recoding techniques (Lan and Anders, 2000). Any disagreements concerning the definition of the classification themes were resolved while the themes were updated. This classification system summarized major research themes emerging in scholarly work and government technology adoption and use over in the past ten years (Zins, 2010). Critical phases in government ICT adoption emerged from content analysis of 360 articles addressing adoption of ICT worldwide. The evolution of government ICT adoption practices since 2008 was documented systematically.

Each paper was further scrutinized and regrouped into three categories according to the specific technology type discussed in the paper, the governance domain the Innovative ICT was applied to, and the specific issue addressed.

Fourth, Similar content analysis and preliminary statistical analysis on 7305 articles included in the E-government Reference Library (Version 13.5) were performed to triangulate finding from qualitative content analysis. The E-Government Reference Library (Version 13.5) is reported as one of the most comprehensive e-libraries of scholarly work about ICT use and public administration. It contains all 9901 articles published in core e-government conference proceedings and journals in the time period of 1981–2017 (Scholl, 2017). Out of them 7305 papers were published in the last decade and served as our basis for analysis in this step.

Based on both the content analysis and the preliminary statistical analysis, eight leading types of innovative ICT adopted in different governance domains in the last decade were finally identified (see Figure 1 for the result). 2105 articles and their citations investigating issues and challenges government and the society is facing when adopting and using these technology were downloaded and systematically recorded.

Finally, a preliminary citation and co-citation analysis was conducted. Five top-cited articles among the 2105 papers were selected before a co-citation analysis was done. Five leading journals and conference proceedings published the biggest number of highly-impacted papers were also examined. Analyses were performed to identify the co-citation patterns across different publication venues. Details of the co-citation analysis can be found in details of our findings.

### 3. Innovative ICT and Advances of Governance (2008-2017)

3.1 What ICT have been adopted in the past decade?

The adoption of ICT in the public sector started as early as in the 1970s. A preliminary analysis of the 7305 articles included in the E-Government Reference Library and their citations demonstrated that researchers paid close attention to the interaction between the newest generation of ICT and governance evolution in the past decade. As can be observed from Figure 1, the last ten years witnessed an acceleration of publications in the e-government discipline. Scholars focused on eight major types of the most innovative ICT – social media, cloud computing, artificial intelligence, big data, data analytics[3], IoT, machine learning, and open data -- and how the use of these technologies interacted with governance practices (Gandomi and Haider, 2015).

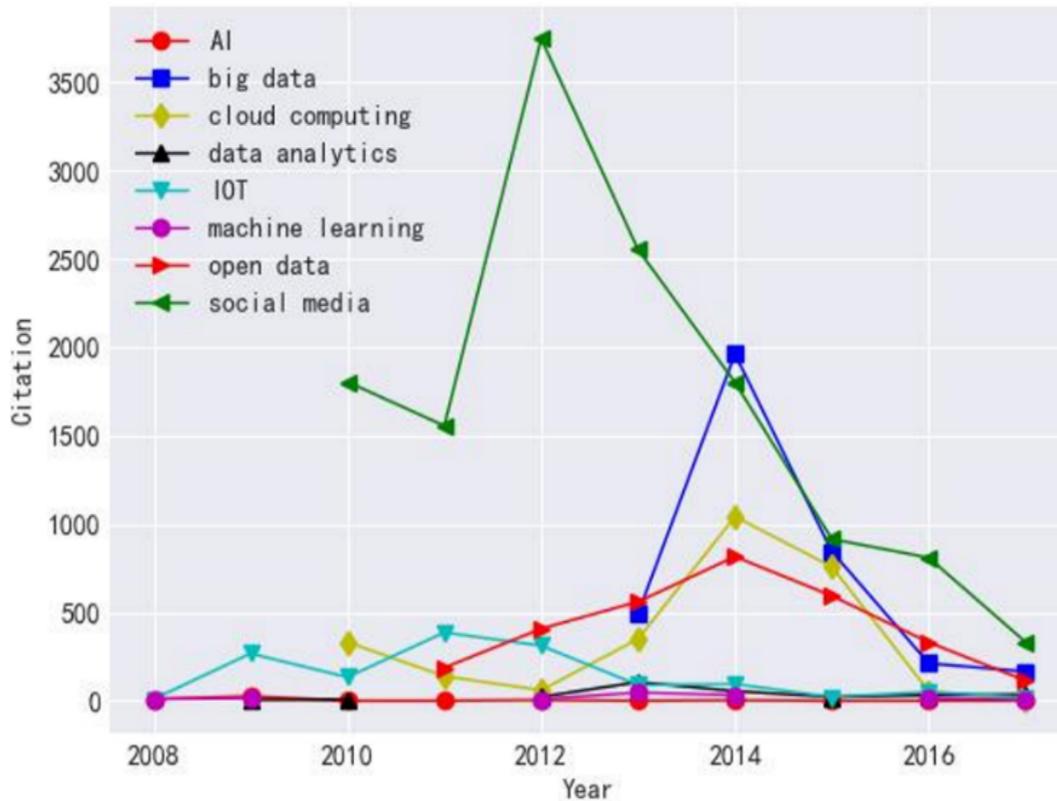

Figure 1 Citations of research on innovative ICT use in governance in the last decade

A closer look at Figure 1 indicates that while social media use for governance purpose is still among top concerns of policy makers and scholars, the research interest on it is drastically descending. On one hand, multiple research published before 2012 emphasizing on social media's challenges to current governance worldwide were among top cited pieces (Bertot et al, 2010; Bertot, Jaeger and Hansen, 2012). This may originate from the power social media has demonstrated in political mobilization and public opinion solicitation. On the other, its potential as an informal data accumulator for public opinion is also attracting wide attentions among scholars promoting open government, open data and civic engagement (Linders, 2010; Lee and Hoon, 2012). However, to successfully unleash the full potential of social media, public sectors are still facing multiple layers of challenges. These challenges include lawmaking and strategy realignment, participatory decision making, procedure design, capability enhancement and crowd co-production in additional to technical proficiency and data analytics capability development (Mergers, 2013; Rodrigo and Ramon, 2012).

---

[3] In the analysis we differentiate between big data and data analytics citing Gandomi and Haider (2015)(p141). Paper labeled with big data are articles discussing the interactions between data management techniques and governance affairs. Data management techniques here include acquisition and recording of data, Extraction, cleaning and annotation of data, Integration aggregation and representation of data. Data analytics here specifically refer to data modelling and analysis, and the interpretation and inference process.

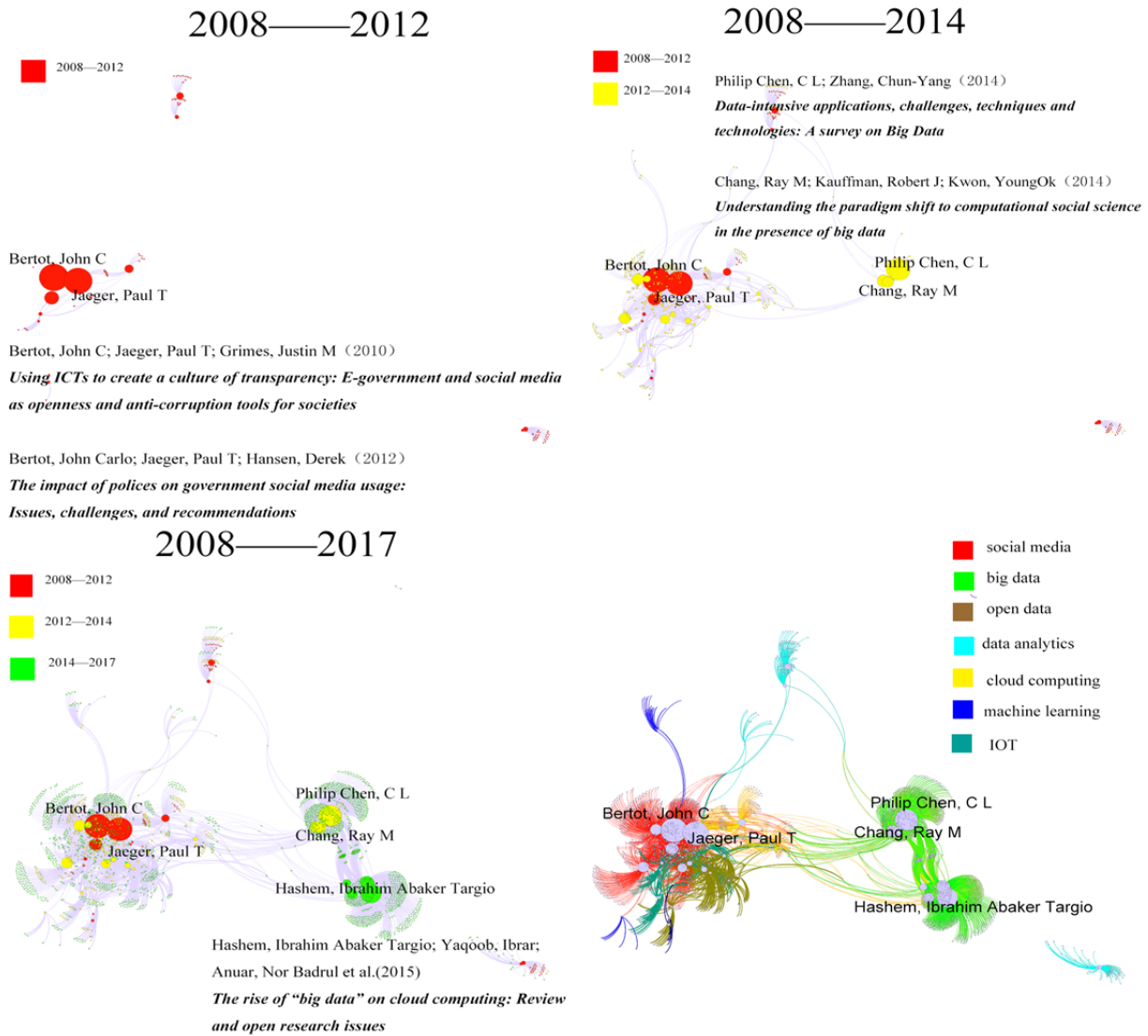

Figure 2 Top five cited articles and their co-citations

It is not surprising to see big data, cloud computing and open data are becoming the next front of innovative technology adoption in governance (Lourenco, 2015;Ohemeng and Adarkwa, 2015; Zuiderwijk et al, 2014; Kalampokis Tambouris and Tarabanis, 2011). Our analysis of the top five cited articles of the E-Government Reference Library (Version 13.5)(Figure 2) published between 2008 and 2017 suggests that there is a clear shift from qualitative theorizing methodologically in the last three years (Romijn and Cunningham, 2017;Palmirani and Girardi, 2016;Koussouris et al, 2015). Many scholars start to reflect on how to utilize huge volume of data available to inform governance and decision making with the quick development of open data and data analytics techniques (Chen and Zhang, 2014; Chang Kauffman and Kwon, 2014)

Cloud computing is in its early stage being adopted as data sharing and service provision platform (Paquette, Jaeger and Wilson, 2010). The academic interests are currently focusing on the technical feasibility, maturity assessment and security risks associated with its adoption (Shin, 2014; Zwattendorf et al, 2013; Khan et al, 2011). For those government in Europe and the US having adopted cloud computing in daily public operations, they are collaborating actively with scholars to understand the challenges and benefits cloud-based platform could introduce to government public interactions (Lian, 2015; Maslina Abawajy and Chowhury, 2013;Knapp Denney and Barner, 2011).

Interestingly, the recent revolution in artificial intelligence and machine learning is presenting potentials of utilizing open and social data for better governance purposes (Teufl, Payer and Parycek, 2009). However, few research on this might implicate that scholars in the e-government discipline are still waiting to observe its interactions with and influence on current governance practices (Moosa and Alsaffar, 2008). Government as traditionally followers instead of pioneers in the innovation diffusion process are still struggling with issues such as secure adoption and technical proficiency to safely manage such new techniques (Liu and Yuan, 2015; Rogers, 1998). Only a few government offices, most law enforcement globally have adopted artificial intelligence and machine learning in its daily operations (Ku and Leroy, 2015).

3.2 How did scholars investigate interactions between innovative ICT and governance core to Internet Plus Government after 2008?

In Figure 3, our citation analysis indicated that five journals and conferences are core for critical finding sharing on interactions between Innovative technology use and governance in the past ten years. These journals and conferences are *Government Information Quarterly*, *Information Sciences*, *Information systems*, *Decision Support Systems*, and *The CIRP Conference on Industrial Product-Service Systems (Figure 3)*.

Different from *Government Information Quarterly* which has established its reputation among e-government researchers, *Decision Support System* was traditionally most sought for by computer scientists. *Information Sciences* and *Information Systems* were designed for researchers in information engineering and intelligent systems. They published articles concerning the design and implementation of languages, data models, process models, algorithms, software and hardware for information systems (Naumann, Shasha and Vossen, 2017). Co-citations and mutual references among work published on these journals and conference proceedings have almost tripled in the past three years (Figure 3).

From Figure 4, we can found that while *Government Information Quarterly* still publish the biggest number of academic research on the use of social media, cloud computing and IoT in governance, other top journals and conferences publish more on the impact of big data and data analytics on governance. At the same time, explorations on machine learning, open data and data analytics are creating more opportunities for researchers from multiple disciplines to collaborate (Kabanaugh et.al, 2012; Effing, Van Hillegersberg, and Huibers, 2011). With the fast development of machine learning and artificial intelligence in the last three years, the interdisciplinary nature featuring close collaborations between data scientists and social scientists is becoming more apparent as evidenced in Figure 4.

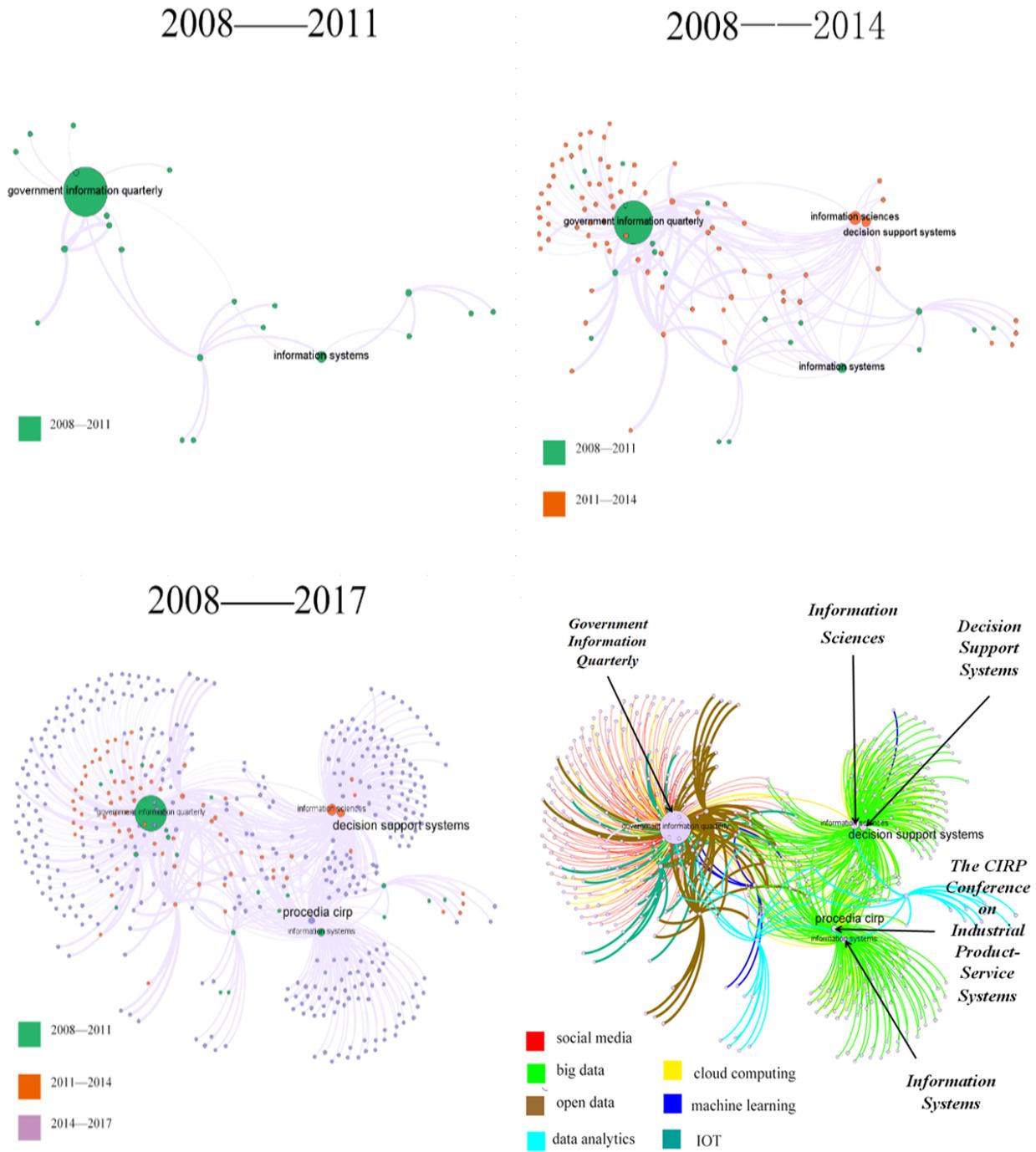

Figure 3 Top publication venues for relevant research

Scholars from different disciplines are also innovating on approaches to decipher the complexity and complication of adopting multiple technology in governance at almost the same time (Ibrahim and Targio, 2015). Our analysis of the top five cited articles in the past decade in the E-gov Reference library substantiates this conclusion. Methods these researchers employed are no longer limited to traditional case study, literature review and quantitative methods (Smith, 2014;Effing, Van Hillegersberg and Huibers, 2011;). Different from research conducted before 2008, more and more scholars from different disciplines start to collaborate with data scientists specializing in data analytics and machine learning algorithms (Piscopo, Siebes and Hardman, 2017; Ku and Leroy, 2015) (Figure 3). More computer scientists and mathematicians are employing machine learning algorithms and data analytical methods to help broaden the perspective (Correa et al, 2014; Patterson et al, 2013).

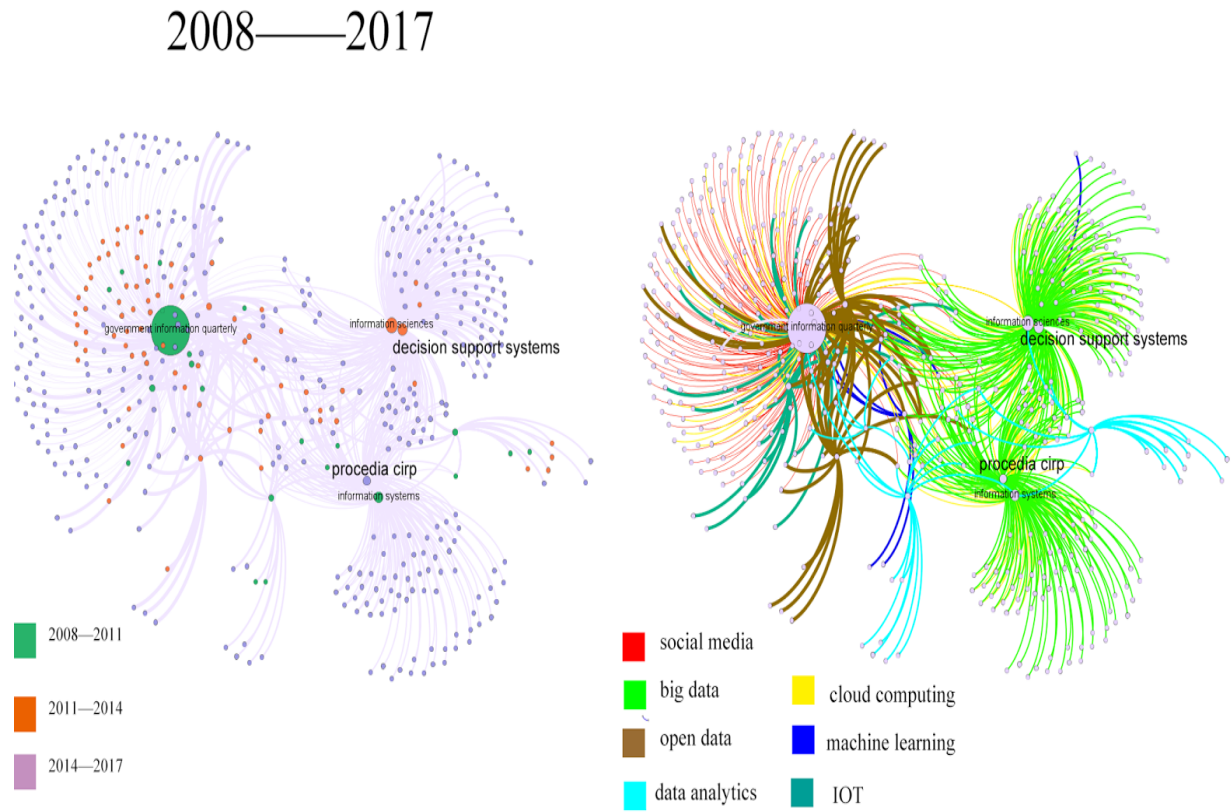

Figure 4  interactions among top publication channels

3.3 What and how governance areas have been impacted?

One everlasting question that continuously haunts the e-government academia is whether and how innovative ICT can actually impact governance practices (Dawes, 2008). Our co-citation analysis suggest that the current research agenda can be divided into three dimensions: technology adoption, data cleansing and use, and issues and challenges emerging in governances as displayed in Figure 5.

When assessing challenges to e-governance a decade ago, Dawes (2008) concluded that the most progress made by then was in enhanced public services and improved management. A variety of innovative technology have since been added to the existing government information infrastructure to support precise service provision and internal management improvement (Lourenco, 2015; Ohemeng and Adarkwa, 2015; Zuiderwijk and Janssen, 2015). Government worldwide have adopted cloud computing platforms, promoted open data initiatives and encourage officials to interact with citizens more frequently and systematically on social media. Figure 5 also presented core governance areas that have witnessed influences from adoptions of multiple types of innovative ICT.

Dawes (2008) pointed out e-democracy, including civil society engagement, and public consultation and political discourse received little attention. This tendency has been changed slowly in the past decade with the wide use of social media. Research contributions include exploration of policy initiatives and strategy design to incorporate public opinion mined from social media. Studies have also attempted to provide theoretical guidance for government to employ cloud computing and machine learning to support citizen engagement and public consultation (Bertot, Jaeger and Hansen,2012). However, explorations as such was still on a more general level of discussion and less on actual practices (Bertot, Jaeger and Grimes, 2010). Very few government have implemented formal procedures for political discourse encouraging public participation in decision and policy making online or offline.

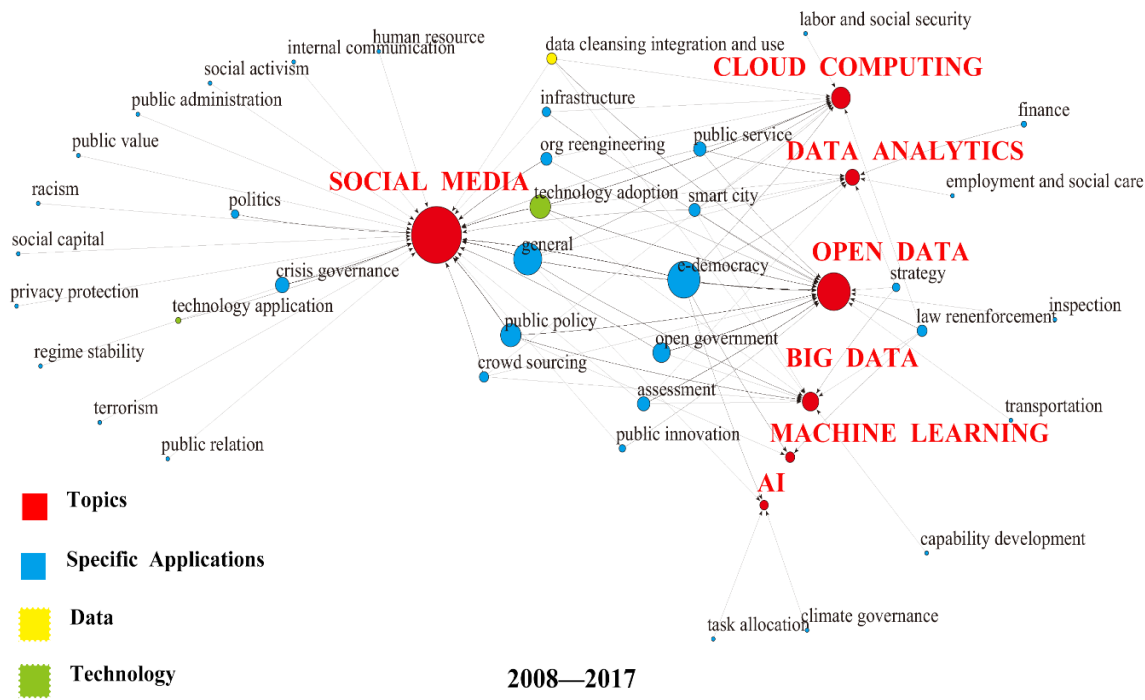

Figure 5 the use of innovative ICT in different governance domains

Open data initiatives promoted by both authoritarian as well as liberal government located on different continents truly created pressure for bureaucrats to share data with the public. This in the longer term might shed light on how to build a more transparent political environment worldwide. The cooperation between data scientists and e-government researchers also provided innovative approaches used to interpret and utilize open and social media data to support transparency. However, currently both government offices and citizens are still struggling with poor data quality and lack of efficient techniques in processing huge amount of data in a real time manner. Not to mention issues and challenges such as privacy infringement associated with integrating social media and open data in business processes and capability development is still daunting for both scholars and practitioners.

Concurrently emerging is a new research direction: the exploration on complications and complexity of adopting multiple types of innovative technology in the bureaucratic governance process. On the one hand, social media obviously have been widely employed in different government sectors. Its significance has been accepted by both the general public and practitioners (Bertot, Jaeger and Glaisyer, 2010). A few research teams have already devoted themselves to investigate how the use of social media, the adoption of cloud computing platforms and open data initiatives can be integrated to support better decision making(Hashem et al, 2015; Chang, Kauffman and Kwon, 2014; Chen and Zhang,2014). On the other, very few research has been conducted to systematically diagnose and predict the complication and risks a huge and complex information infrastructure might introduce to practitioners. Since the millennium change network invasive computer viruses prompted awareness, education, and monitoring efforts and led to a market for new products to protect information and systems from hackers and other threats.

However, research on these security measures are mostly technical. They focus only on analysis of the potential risk associated with one single technology without relating it to the broad government information infrastructure system (Conradie and Choennin, 2014; Sebastien et al, 2013). Discussions of the life cycle of technology adoption, implementation and use are not systematically paralleled with the evolution of bureaucratic operation. Thus government IT workforce and civil servants constantly found themselves debating. The adoption of a new technology without having the full picture of challenges associate with a complicated and multiple-layered information and data structure demands for more comprehensive investigation.

3.4 What is the relationship between the numbers of publications using different ICT technologies and countries?

We use Figures 6-9 to demonstrate the use of different types of innovative ICT in different countries worldwide. The deeper the color is, the more influential is the specific type of ICT adopted in that country.

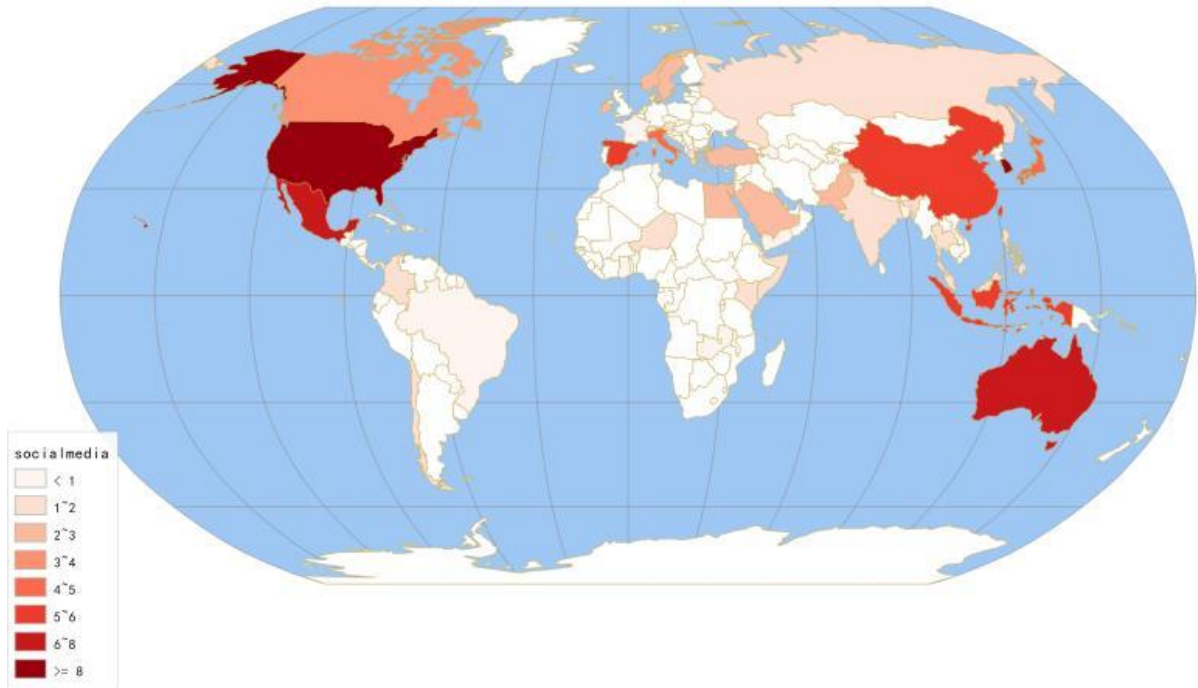

Figure 6 the numbers of ICT papers using social media in different countries

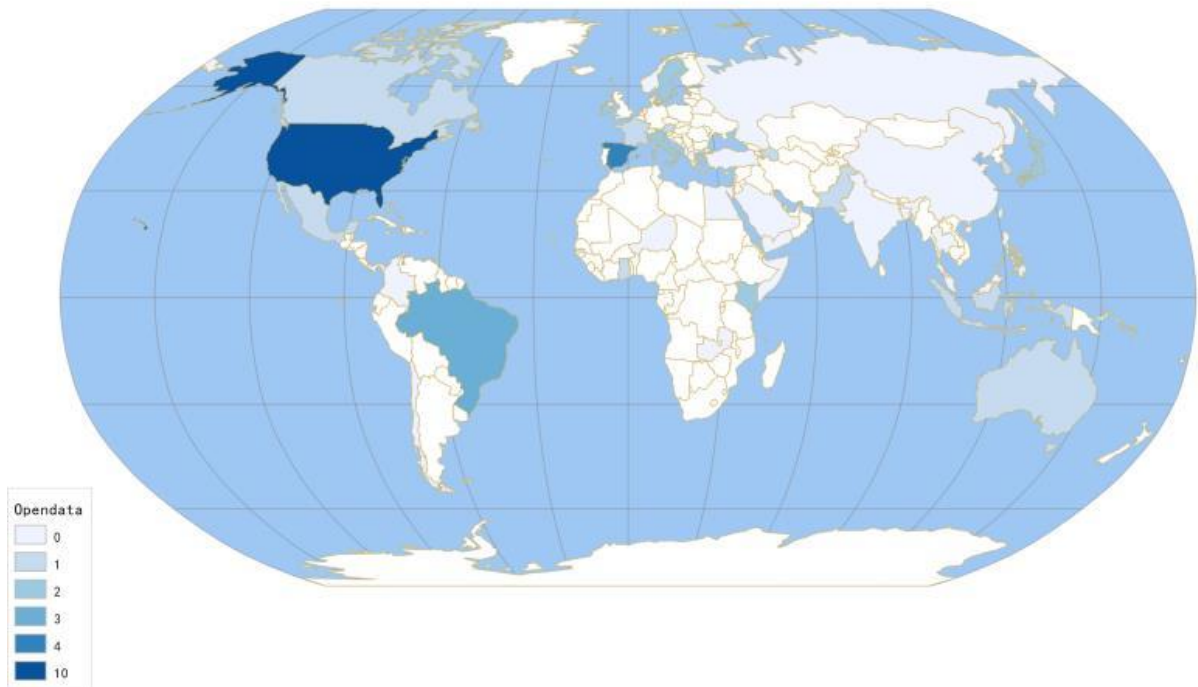

Figure 7 the numbers of ICT papers using open data in different countries

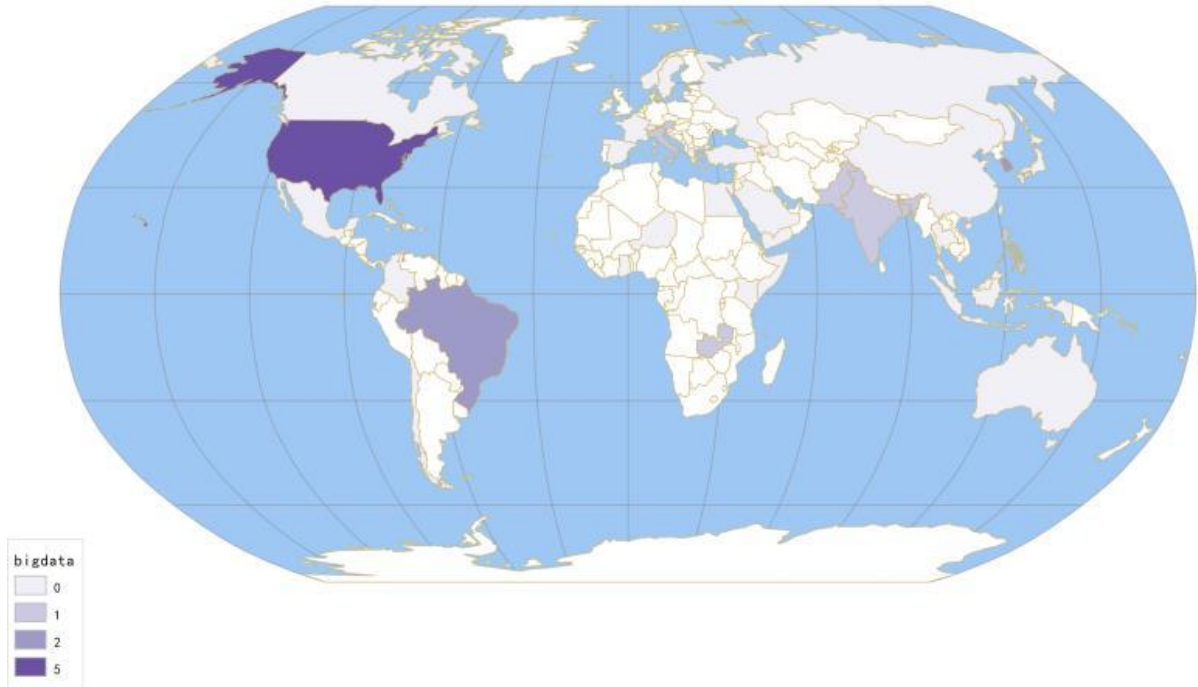

Figure 8 the numbers of ICT papers using big data in different countries

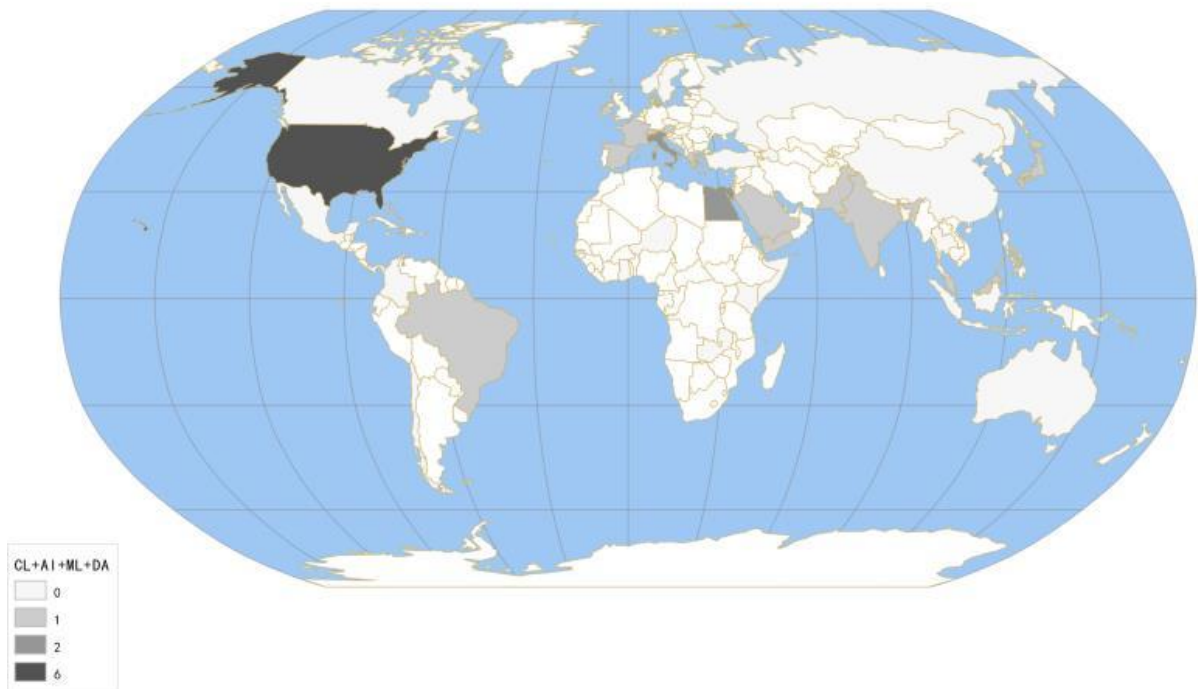

Figure 9 the numbers of ICT papers using CL, AI, ML and DA in different countries

## 4. Concluding remarks: Where are we heading?

From the analysis, we can see that practitioners aim to incorporate the depth of innovative networking technology in governance in the past ten years.

However, governments adopting the Internet Plus Government initiative also hope to develop a more sustainable model of social development. They hope to utilize Web 2.0 as the fundamental facility and implementation tool throughout the governance processes and activities. Any government hoping to solely focus on economic development and bypass the political transformation might actually become very brittle in the longer term.

Furthermore, the use of data analytics and machine learning entitled in Internet Plus Government to support decision and policy making may present both challenges and opportunities to government and academia. Data modelling and prediction techniques make it more convenient for government to mine the internet to understand the true needs of citizens. It also equips the scholars with prediction tools to go beyond the descriptive nature limited by certain qualitative methods. But it also provides challenges to government and researchers who may be lack of such capability (Chatfield Reddick and Al-Zubaidi, 2015). What is even more daunting is the risk embedded in certain government's employing similar technology for monitoring and witch-hunting purpose. How to balance between benefits associated with the new wave of data-driven innovations in the government and citizen right protection is definitely among top issues to be addressed in the coming seasons.

Our review highlighted the research directions and questions remain to be addressed in the domain of e-democracy. Members of the e-governance community still need to systematically observe and clearly delineate the process where technology and governance interact and mutually shape each other. Government officials also need to be more tolerant when it comes to the profound effects generated by new technology adoption and use in different governance domains. It is only through the transparent collaboration among the government, the academia and the citizens can the full potential of the newest wave of networking technology be unleashed for better governance.

Zuiderwijk, A., Helbig, N., Gil-García, J. R., & Janssen, M. (2014). Special Issue on Innovation through Open Data: Guest Editors' Introduction. Journal of theoretical and applied electronic commerce research, 9(2), i-xiii.

ZWATTENDORFER, B., HILLEBOLD, C., AND TEUFL, P. Secure and privacy-preserving proxy voting system. ICEBE '13, pp. 472–477